\documentclass{article}

\usepackage{PRIMEarxiv}

\usepackage[utf8]{inputenc} 
\usepackage[T1]{fontenc}    
\usepackage{hyperref}       
\usepackage{url}            
\usepackage{booktabs}       
\usepackage{amsmath,amsfonts}       
\usepackage{nicefrac}       
\usepackage{microtype}      
\usepackage{lipsum}
\usepackage{fancyhdr}       
\usepackage{graphicx}       
\usepackage{subcaption}
\usepackage{mdframed}
\usepackage{makecell}
\usepackage{multirow}
\usepackage[table,xcdraw,svgnames]{xcolor}
\usepackage{threeparttable}
\graphicspath{{media/}}     

\title{HKT-SmartAudit: Distilling Lightweight Models \\ for Smart Contract Auditing
}

\author{
  Zhiyuan Wei \\
  School of Computer Science and Technology \\
  Beijing Institute of Technology \\
  Beijing, China\\
  \texttt{weizhiyuan@bit.eud.cn} \\
   \And
  Jing Sun \\
  School of Computer Science \\
  University of Auckland \\
  Auckland, New Zealand\\
  \And
  Zijian Zhang, Xianhao Zhang \\
  School of Cyberspace Science and Technology \\
  Beijing Institute of Technology \\
  Beijing, China\\
  \And
  Zhe Hou \\
  School of Information and Communication Technology \\
  Griffith Univerity \\
  Nathan, Australia\\
}

\begin{document}
\maketitle

\begin{abstract}
The rapid growth of blockchain technology has driven the widespread adoption of smart contracts; however, their inherent vulnerabilities have led to significant financial losses. Traditional auditing methods, while essential, struggle to keep pace with the increasing complexity and scale of smart contracts. Large Language Models (LLMs) offer promising capabilities for automating vulnerability detection, but their adoption is often limited by high computational costs. Although prior work has explored leveraging large models through agents or workflows, relatively little attention has been given to improving the performance of smaller, fine-tuned models—a critical factor for achieving both efficiency and data privacy. In this paper, we introduce HKT-SmartAudit, a framework for developing lightweight models optimized for smart contract auditing. It features a multi-stage knowledge distillation pipeline that integrates classical distillation, external domain knowledge, and reward-guided learning to transfer high-quality insights from large teacher models. A single-task learning strategy is employed to train compact student models that maintain high accuracy and robustness while significantly reducing computational overhead. Experimental results show that our distilled models outperform both commercial tools and larger models in detecting complex vulnerabilities and logical flaws, offering a practical, secure, and scalable solution for smart contract auditing. The source code is available at Github repository\footnote{https://github.com/LLMSmartAudit/FTSmartAudit}.
\end{abstract}


\section{Introduction}
{Recent} advancements in distributed ledger technologies have precipitated significant progress in the domain of smart contracts, with notable implications for applications such as cybersecurity and decentralized finance \cite{xing2025zero}. Smart contracts, defined as self-executing programmatic agreements with predefined conditions established by participating entities, have emerged as pivotal innovations in decentralized systems. While smart contracts offer significant advantages in terms of automation and trustless transactions, they are also prone to vulnerabilities and bugs that can lead to severe financial and operational consequences. Over the past half-decade, the ecosystem has witnessed a series of high-profile security breaches and exploitations targeting smart contracts, such as the DAO attack~\cite{chen2020survey}. Just 2024 alone, total financial losses exceeded \textbf{2.6 billion} USD across \textbf{192 incidents}, resulting from smart contract-level vulnerabilities\footnote{https://getfailsafe.com/failsafe-web3-security-report-2025/}.
Consequently, ensuring the security of smart contracts remains a complex challenge.

Large Language Models (LLMs), including BERT, T5, and GPT, have demonstrated strong potential in automating vulnerability detection by extracting key features and providing accurate predictions \cite{li2024enhancing, zhang2024pydex, raffel2020exploring}. For example, Zhang et al. \cite{zhang2024prompt} highlighted GPT’s effectiveness in detecting vulnerabilities in C/C++ and Java, often outperforming traditional methods. 
Despite their impressive capabilities, the practical deployment of LLMs is often constrained by substantial computational demands. Serving a single model with 175 billion parameters at 16-bit (FP16/BF16) precision requires at least \textbf{350 GB} of GPU memory and specialized infrastructure \cite{zheng2022alpa}. Larger models such as DeepSeek yield superior performance has fueled the development of models with a total of \textbf{671 billion} parameters. However, such requirements place these models far beyond the reach of most users, particularly in scenarios where low-latency performance is essential. 
The practical deployment of commercial proprietary LLMs often incurs significant computational and financial costs, especially at enterprise scale or with frequent usage. In addition, relying on third-party LLM providers introduces serious security and privacy risks, as sensitive data must be transmitted to external servers—such as source code or system configurations—must be transmitted to external servers. Excessively large models may also carry redundant or irrelevant knowledge, impairing performance on domain-specific tasks \cite{he2023large, zhao2023survey}. 

In contrast, open-source, smaller LLMs provide a more efficient and secure alternative, supporting local deployment, reduced infrastructure costs, and greater control over sensitive data. This makes them particularly suitable for domain-specific applications like smart contract auditing, where tasks are narrowly scoped and require precise reasoning. While prior research has primarily focused on leveraging existing LLMs via agents or workflows, relatively little attention has been given to enhancing the models themselves \cite{liu2024propertygpt, sun2024gptscan}. The development of lightweight, specialized models remains underexplored, despite their potential for high performance in targeted auditing tasks. 
Moreover, although models like DeepSeek and LLaMA series are labeled as open-source, they restrict access to training data, offering only model weights or inference capabilities. This lack of transparency poses substantial challenges for fine-tuning and adapting models to specific domains. The shift toward smaller models is thus driven not only by economic considerations but also by growing demands for secure, transparent, and adaptable AI solutions. 

To address these challenges, we propose \textbf{HKT-SmartAudit} (Hybrid Knowledge Transfer for Smart Contract Audit), a comprehensive framework aimed at enhancing the auditing capabilities of compact language models. This work introduces SOTA, \textbf{task-specialized small} models tailored for smart contract auditing and presents effective methods for extracting and curating high-quality knowledge from both advanced large-scale models and external domain-specific sources. Our key contributions are summarized as follows:

\begin{itemize}
\item \textbf{Hybrid Knowledge Transfer Framework}: We introduce HKT-SmartAudit, a novel framework that transfers knowledge from both advanced large language models and external domain-specific sources. The framework ensures that compact student models effectively absorb this knowledge through high-quality, task-specific training datasets and targeted fine-tuning techniques, significantly improving their auditing performance.
\item \textbf{Iterative Learning with Reward-Guided Refinement}: We develop a reward model to facilitate continual learning and refinement of specialized models. By maintaining an up-to-date repository of real-world audited smart contracts, our system enables ongoing auditing by student models. Their outputs are evaluated using the reward model, and high-quality results are used to iteratively improve the models’ accuracy and robustness.
\item \textbf{Improved Vulnerability Detection}: Our specialized models achieve superior performance in detecting both known vulnerabilities and complex, logic-driven security flaws. This enhanced detection capability strengthens the overall reliability and security of smart contracts.
\item \textbf{Model Extensibility and Generalization}: While tailored for smart contract auditing, our framework is inherently extensible and can be applied to other domains that require LLM-based analysis. Potential applications include code quality assessment, regulatory compliance checking, and broader security-related tasks.
\end{itemize}


\section{Backgrounds}
\label{sec:backgrounds}
\subsection{Smart Contract Security}

Smart contracts have emerged as a transformative force in the digital realm, giving rise to a wide range of compelling applications. Recent surveys \cite{wei2023survey, zhou2023sok} indicate a rapid increase in the number of smart contracts over the past five years. DeFi, the most important application of smart contracts, has seen a significant rise in popularity, with its peak total value locked (TVL) reaching 179 billion in USD on 9 November 2021\footnote{https://defillama.com/}. However, the substantial asset values associated with smart contracts also attract numerous potential malicious actors. Smart contracts have been plagued by several high-profile vulnerabilities and exploits. Zhou et al. \cite{zhou2023sok} document that smart contracts have suffered from countless high-profile attacks, resulting in losses exceeding 3.24 billion in USD from April 2018 to April 2022. 

Unlike traditional software, smart contracts are more prone to having vulnerabilities permanently embedded within their code. Thus, many security researchers try to find analysis tools to automatically detect and analyze smart contracts before they are deployed. They employ advanced techniques such as formal verification, symbolic execution, fuzzing, and intermediate representation (IR) to enhance their effectiveness \cite{chen2020survey, tolmach2021survey}. While essential, these traditional auditing methods are often time-consuming and may not scale well with the growing complexity and number of smart contracts.
For instance, tools relying on formal verification are adept at ensuring contracts adhere to specified requirements but may fall short in detecting security flaws like reentrancy or gas limit. 

\subsection{LLMs for Vulnerability Prediction}
LLMs are termed ``large'' due to their extensive number of parameters, which empower them to comprehend and generate human language with remarkable coherence and contextual appropriateness. Natural language processing, image generation, code, and mathematical problem-solving are on the topic of target domains. In the realm of code processing, LLMs have shown considerable advancement since the pioneering work of Codex \cite{chen2021evaluating}. This progress has led to the development of commercial products like GitHub Copilot \cite{barke2023grounded} and open-source code models such as StarCoder and Codellama \cite{roziere2023code}. 

LLMs have also achieved excellent performance on specific downstream tasks, such as code analysis, vulnerability detection and code upgrading \cite{ding2024cycle}. Chen et al. \cite{chen2023diversevul} have proven that LLMs (GPT-2, T5), trained with a high-quality dataset consisting of 18,945 vulnerable functions about C/C++, outperform other machine learning methods, such as Graph Neural Networks, in vulnerability detection. Additionally, fine-tuned models such as CodeT5 and NatGen significantly improve the performance on vulnerability detection task. SOTA commercial products, such as GPT-4, offer promising solutions to augment the smart contract auditing process \cite{david2023you, chen2023chatgpt}. 
By leveraging their code comprehension and generation capabilities, these models can identify specific vulnerabilities, verify compliance, and check for logical correctness. Their effectiveness is further enhanced through advanced prompting techniques like chain-of-thought (CoT) or few-shot.

\subsection{Knowledge Distillation}
Knowledge distillation is a model compression technique where a smaller, simpler model ("student") learns from a larger, more complex model, known as the "teacher."
This process involves training the student model using the teacher model's final predictions (hard targets) or soft output distributions (soft labels), capturing richer knowledge \cite{beyer2022knowledge}. This method enables the deployment of lightweight student models on resource-limited devices, such as embedded systems and home devices. Central to knowledge distillation is the student-teacher architecture, which ensures effective knowledge transfer from a computationally intensive teacher model to a compact student model optimized for environments with limited resources \cite{gou2021knowledge}. This architecture allows the student model to inherit critical insights from the teacher while significantly reducing computational demands.

Knowledge transferred from teacher to student includes soft labels, intermediate layer representations, and structural relationships across different layers and data instances \cite{acharya2024survey}. The effectiveness of knowledge distillation greatly depends on the quality of the teacher model’s training, particularly the accuracy of data annotations. Additionally, selecting an appropriate distillation algorithm is crucial. Prominent distillation approaches include soft target distillation, adversarial learning, cross-model distillation, multi-teacher distillation, and graph-based distillation. In practical application, compared to retraining all of a model's parameters from scratch, training a student model is more efficient and can achieve significant improvements with fewer resources. For example, CodeLlama model \cite{roziere2023code} was created by fine-tuning Llama 2 on a mix of proprietary instruction data. 




\begin{figure}[t]
    \centering
    \includegraphics[width=5.5in]{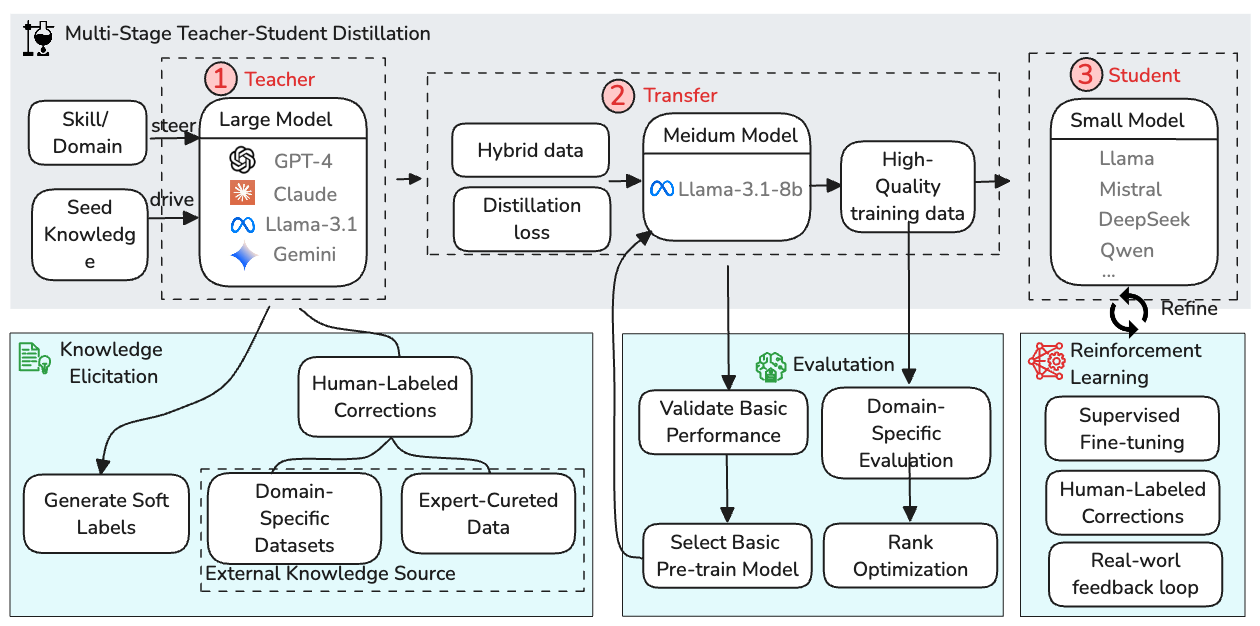}
    \caption{An overview of HKT-SmartAudit framework.}
    \label{framework}
\end{figure}

\section{Methodology}
\label{sec:methodology}

We introduce HKT-SmartAudit (Hybrid Knowledge Transfer for Smart Contract Audit), a novel framework that combines classical knowledge distillation, external domain knowledge, and reinforcement learning (RL)-enabled fine-tuning to develop a specialized model for smart contract auditing. The proposed approach retains the generalization capabilities and rich representations of a large teacher model while incorporating domain-specific insights derived from curated datasets and expert feedback. As illustrated in Fig.~\ref{framework}, the framework comprises four key components: Multi-Stage Teacher-Student Distillation, Knowledge Elicitation, Reinforcement Learning, and Evaluation -- each of which is described in detail below.

\subsection{Multi-Stage Teacher-Student Distillation}
\label{sec:teacher_student}
At the core of HKT-SmartAudit is a multi-stage teacher-student paradigm that extends beyond conventional distillation methods by incorporating a progressive transfer of knowledge. This paradigm consists of three components: the teacher model, the student model, and the knowledge transfer process.

\paragraph{Powerful teacher model}
We begin by selecting a robust teacher model that demonstrates strong performance on the target task. The teacher’s soft outputs serve as informative and reliable training targets. Before distillation, the teacher is evaluated on a validation set to ensure its accuracy and capacity to capture the complexities of smart contract auditing. As shown in Figure \ref{fig:various_model_eval}, large parameters models (e.g., GPT-4, Claude) outperform smaller models (e.g., llama3.1-8b, mistral-7b-v0.3), ensuring that the teacher delivers both domain expertise and effective learning guidance.

\paragraph{Efficient Student Model}
The student model is selected based on its efficiency; it must be significantly smaller or faster than the teacher while remaining expressive enough to learn the task accurately. We consider models with smaller transformer architectures, fewer layers, or reduced hidden units—often distilled variants of the teacher’s architecture. There is no strict requirement for the teacher and student to share the same architecture, but the student must retain sufficient representational power to absorb critical knowledge. As illustrated in Fig. \ref{fig:standard_set}, models with approximately 1 billion parameters (e.g., llama3.1-1b, mistral-3b) perform considerably lower than models with 10 or more billion parameters (e.g., llama3.1-8b, mistral-7b-v0.3). Our model training follow an iterative approach to identify the minimal student model size that maintains high performance.

\paragraph{Establishing the Knowledge Transfer Process}
With both teacher and student models in place, the student is trained to mimic or even outperform the teacher using high-quality transferred knowledge. One challenge is ensuring that the student captures the nuanced expertise of the teacher, especially when the student is much smaller. To mitigate this, we introduce an intermediate model to assess the quality of the transferred knowledge. For critical knowledge areas, if the student’s performance regresses, we explicitly use reinforce learning by incorporating additional training examples or adjusting the loss function to focus on those aspects.

\subsection{Knowledge Elicitation}
\label{sec:knowledge_elicitation}

\begin{figure*}[t]
    \centering
    \includegraphics[width=5.5in]{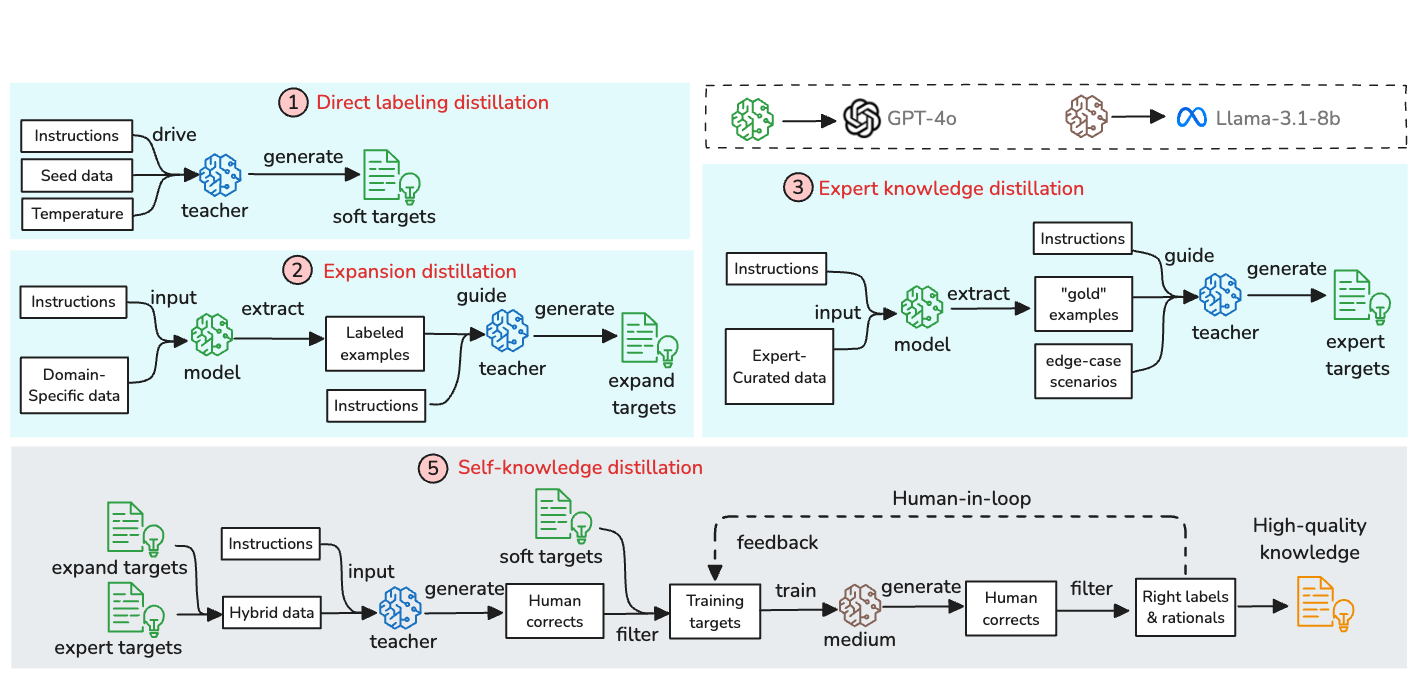}
    \caption{An illustration of knowledge transfer methods from teacher LLMs to small models.} 
    \label{fig:knowledge}
\end{figure*}

To facilitate effective knowledge transfer to small models, we propose four complementary methods: direct labeling distillation, expansion distillation, expert knowledge distillation, and self-knowledge distillation. These approaches are illustrated in Fig.~\ref{fig:knowledge} and described in detail below.

\paragraph{Direct Labeling Distillation} In this step, the teacher model is driven by seed knowledge to obtain the desired generations. Rather than producing a single “correct” answer, the teacher outputs a probability list (i.e., soft labels) that contains detailed insights on plausible alternatives. The temperature parameter $T$ (with $T > 1$ resulting in a “softer” distribution) can be tuned to adjust the spread of these probabilities. For example, a distribution of (0.7, 0.2, 0.1) offers richer guidance than a hard label. By the end of this method, the student model begins approximating the teacher’s behavior.

\paragraph{Expansion Distillation} Solely distilling knowledge from the teacher limits the student to replicating the teacher’s ability. To enrich the student model and address potential deficiencies in the teacher, we incorporate additional domain-specific data. We collect text corpora from various sources (e.g., OpenZeppelin, CVE, SWC, and entethalliance) and use an advanced model (such as GPT-4o) to extract key information from expert-labeled examples. These curated examples then guide teacher model in generating expanded targets, ensuring the student learns domain-specific terminology, styles, and patterns that general teacher might not fully capture.

\paragraph{Expert Knowledge Distillation} In addition to domain-specific corpora, it is critical to incorporate edge-case scenarios and known challenging vulnerabilities (e.g., logical or semantic flaws, zero-day simulations). We curate a small set of “gold” question-answer pairs, edge-case examples, or corrected scenarios verified for accuracy, drawing from live audited reports from security firms such as Code4rena and Sherlock. Training on this high-quality, expert-validated data enables the student to grasp critical reasoning skills and nuanced knowledge that may be absent from the teacher’s general output. As with expansion distillation, these expert examples are processed by an advanced model to ensure consistency in format and representation.

\paragraph{Self-Knowledge Distillation} Evaluating whether the elicited knowledge suits the student is as important as ensuring it aligns with the teacher. Here, we leverage self-knowledge by combining expansion and expert targets as inputs to the teacher model. The teacher’s outputs are then reviewed by human experts, and any errors are corrected to produce accurate labels or answers. These corrected outputs, together with the soft targets, are integrated into the student’s training data. This process not only corrects teacher errors but also reinforces the student’s learning through a human-in-the-loop feedback loop (model proposes → human corrects → model retrains). Over time, this iterative refinement significantly enhances the student’s reliability. Moreover, to ensure the highest quality of distilled knowledge, we select a high-performance small model (llama-3.1-8b-it) as the student.

\subsection{Reinforcement Learning}
\label{sec:reinforcement_learning}

\begin{figure*}[t]
    \centering
    \includegraphics[width=5.4in]{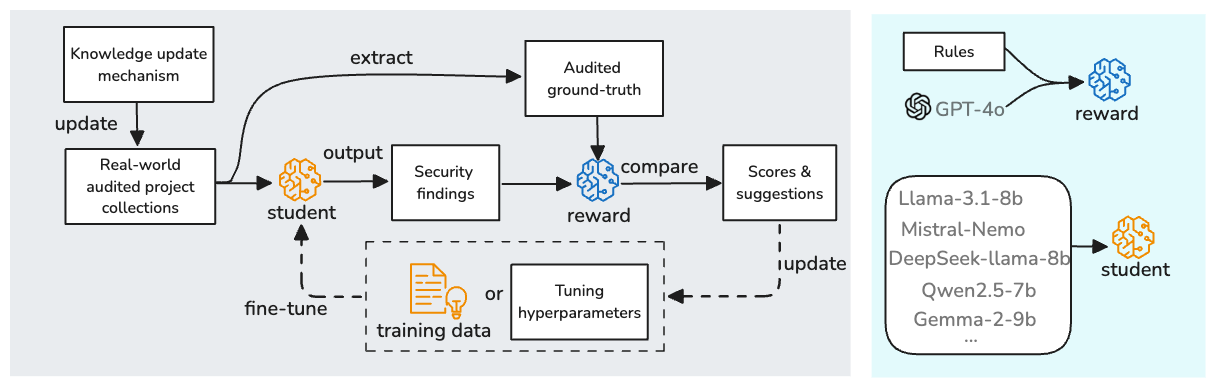}
    \caption{An illustration of reinforcement learning.}
    \label{fig:reinforcement}
\end{figure*}

In Section~\ref{sec:teacher_student}, we selected LLaMA-3.1-8B as the medium-sized model to generate high-quality training knowledge. While this approach is effective for models with similar architectures (in terms of speed and size), it may not generalize well to different architectures. Therefore, further refinement is necessary for long-term adaptability. In our framework, we enhance the student model using reinforcement learning (RL) \cite{ji2024reinforcement} to adapt to real-world feedback.

At this stage, the model is refined through iterative feedback from multiple sources, including human evaluations, environmental signals, and automated assessments provided by a reward model. As illustrated in Fig.~\ref{fig:reinforcement}, the fine-tuning process—guided by additional training data and optimized hyperparameters—enables the student model to improve incrementally while maintaining significantly higher efficiency than the original teacher model. The reinforcement learning process is driven by the following key components:

\begin{itemize}
    \item \textbf{Define a Reward Signal}: First, we define the criteria for desirable behavior in the student model. To achieve this, we use real-world audited project collections as ground truth. The student model processes the smart contracts, while the corresponding audit reports serve as references. A reward model then compares the student’s outputs against these reports, calculates the prediction loss, and generates feedback scores to guide training.
     \item \textbf{Knowledge Update Mechanism}: A web crawler scrapes target platforms (e.g., Code4rena, Immunefi) weekly for newly audited contracts and reports. Then, only reports with clear vulnerability classifications are retained. Finally, new audits are appended to the training corpus, ensuring the reward model and student adapt to the latest security trends.
    \item \textbf{Iterative Feedback Loop:} The RL phase proceeds iteratively. In each iteration, the student model generates outputs that are evaluated by a reward model. Rather than relying solely on human feedback, our reward model combines insights from an advanced model with rule-based criteria. This hybrid approach not only labels security findings as correct or incorrect but also identifies partial matches and discrepancies. The training loop continues until the model’s performance stabilizes and meets predefined quality thresholds on both benchmark and domain-specific test sets.
    \item \textbf{Enforce Domain Constraints:} By carefully shaping the reward function, we can penalize the student model for undesirable behaviors. For example, if the domain requires the model to never violate a safety rule, any output that breaks the rule receives a large negative reward. If certain outputs are preferred—such as more concise explanations—the reward can be adjusted accordingly. Over many iterations, the student model learns policies that respect these constraints in order to maximize its overall reward.
    \item \textbf{Fine-tuning:} For retraining, we employ Supervised Fine-Tuning (SFT) in conjunction with Quantized Low-Rank Adaptation. As RL fine-tuning can occasionally lead to the forgetting of previously acquired supervised knowledge, we continuously monitor both the reward signal and traditional evaluation metrics to ensure consistent performance.
\end{itemize}

\section{Model Selection and Training}
\label{sec:fine-tuning}
This section explores the intricate process of fine-tuning target models specifically for the task of smart contract auditing. 

\subsection{Construction of Training Datasets}
As described in Section \ref{sec:knowledge_elicitation}, to build a robust student model, we require several types of datasets: seed data, soft targets, expanded targets, and expert targets. All generated datasets follow the format of a question–answer pair (contract-findings), where the findings include `labels' and `rationales'.

\begin{itemize}
    \item {Seed Set:} This dataset is drive to prompt the teacher model to generate soft targets. To our knowledge, there are 120 distinct types of vulnerabilities. Accordingly, we collected 120 human-labeled contracts covering these vulnerability types along with their corresponding explanations.
    \item {Soft Targets Set:} This dataset is mainly distilled from the teacher model using seed set. Instead of using one true label (hard label), we ensure that each contract entry has at least two labels (soft label). This set comprises 1,800 entries.
    \item {Expanded Targets Set:} This dataset originates from domain-specific text corpora, such as OpenZeppelin, CVE, SWC, and EntethAlliance. From these resources’ documentation and code repositories, we extracted labeled examples and processed them through the teacher model, resulting in approximately 421 entries.
    \item {Expert Targets Set:} This dataset is sourced from curated vulnerabilities that have been explicitly annotated by smart contract experts, particularly from real-world projects (e.g., Code4rena audited-report). We extracted these audited examples and processed them via the teacher model, yielding about 520 entries.
\end{itemize}

These datasets are merged into a single initial training set. After several iterations of the self-knowledge distillation process, we obtained a final training set, \textbf{HKT-vul}, containing 2,434 entries. During this process, we observed that the student model was unable to absorb all of the provided knowledge effectively, which led us to discard certain entries. As a result, the final training set is smaller than the initially aggregated 2,741 entries.

For comparison, we also introduce secure contracts into our training process. To facilitate this, we created a secure dataset (\textbf{HKT-sec}) by removing vulnerable contracts from HKT-vul. We then constructed a mixed training set (\textbf{HKT-mix}) that includes both vulnerable and secure entries. Table~\ref{tab:dataset_statistics} summarizes key statistics for our training datasets, presenting the number of contracts categorized by complexity (Low, Medium, High) for both the vulnerable dataset (HKT-vul) and the mixed dataset (HKT-mix).

\begin{table*}[h]
  \centering
  \footnotesize
  \caption{Dataset Statistics for HKT-vul and HKT-mix}
  \label{tab:dataset_statistics}
  \begin{tabular}{llcc}
    \toprule
    \textbf{Complexity Level} & \textbf{Characteristics} & \textbf{HKT-vul } & \textbf{HKT-mix } \\
    \midrule
    \textbf{Low Complexity} & 
    \begin{minipage}{0.3\textwidth}
      \begin{itemize}
        \item Less than 500 tokens.
        \item Simple logic (e.g., basic transfers, ownership management).
        \item Few or no loops, conditionals.
      \end{itemize}
    \end{minipage} &
    1510 & 3020 \\
    \midrule
    \textbf{Medium Complexity} & 
    \begin{minipage}{0.3\textwidth}
      \begin{itemize}
        \item 500 to 2,000 tokens.
        \item Moderate logic (e.g., multi-step transactions, basic governance, or staking mechanisms).
        \item Some external calls or integrations (e.g., oracles, DeFi protocols).
      \end{itemize}
    \end{minipage} &
    711 & 1422 \\
    \midrule
    \textbf{High Complexity} & 
    \begin{minipage}{0.3\textwidth}
      \begin{itemize}
        \item 2,000+ tokens.
        \item Complex logic (e.g., multi-contract systems, advanced governance, or financial instruments).
        \item Heavy use of external calls, libraries, and dependencies.
      \end{itemize}
    \end{minipage} &
    213 & 426 \\
    \midrule
    \textbf{Total Contracts} & & \textbf{2,434} & \textbf{4,868} \\
    \bottomrule
  \end{tabular}
\end{table*}

\subsection{Teacher-Student Model Selection and Evaluation}
\label{sec:teacher_student_model}

\begin{figure}[htp]
    \centering
    \begin{subfigure}[b]{0.48\textwidth}
        \centering
        \includegraphics[width=\linewidth]{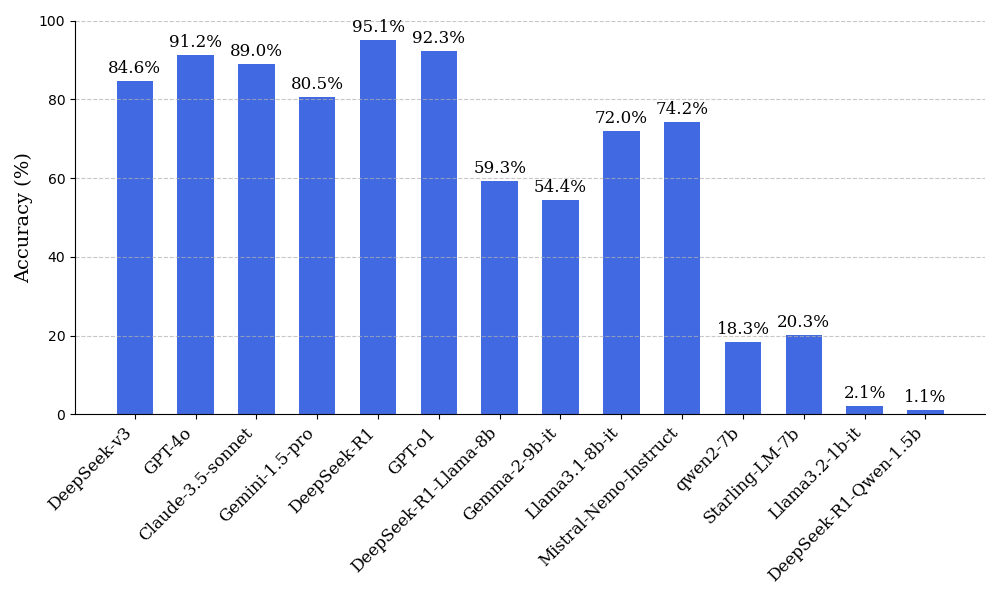}
        \caption{Standard Set}
        \label{fig:standard_set}
    \end{subfigure}
    \hfill
    \begin{subfigure}[b]{0.48\textwidth}
        \centering
        \includegraphics[width=\linewidth]{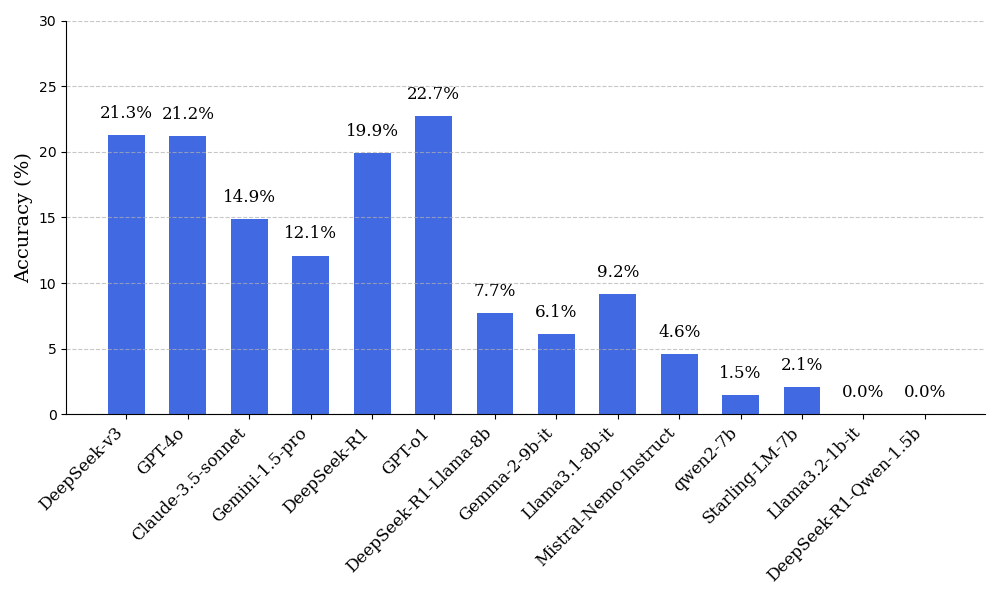}
        \caption{Real-world Project Set}
        \label{fig:realworld_set}
    \end{subfigure}
    \caption{Performance Comparison on Two Sets}
    \label{fig:various_model_eval}
\end{figure}

Given the wide range of available language models, selecting appropriate teacher (larger, advanced) and student (smaller, efficient) models is crucial for this study. We began by performing a filtering step to identify suitable candidates. For the teacher models, we exclusively considered advanced models accessible via API calls.

The selection criteria for the student models—specifically tailored for smart contract auditing—are defined as follows:

\begin{itemize}
  \item \textit{Solidity Comprehension}: Ability to parse and comprehend Solidity, the primary language for Ethereum contracts.
  \item \textit{Code-related Task Performance}: Proficiency in tasks such as code completion, vulnerability detection, and code summarization.
  \item \textit{Semantic Understanding}: Capacity to grasp the semantic meaning of code beyond mere syntax.
  \item \textit{Model Type}: Focus on completion models such as GPT and Llama series, which offering prediction for the most likely next words or tokens.
  \item \textit{Model Size}: Preference for models not exceed 10 billion parameters to demonstrate the efficacy of smaller models and reduce computational requirements.
  \item \textit{Performance and Flexibility}: Emphasis on models that offer high performance and can be fine-tuned for specific smart contract auditing tasks.
  \item \textit{Community and Support}: Prioritizing models with active community support and comprehensive documentation (e.g., those available on Hugging Face) for easier integration and troubleshooting.
\end{itemize}

Based on these criteria, we selected the following models for evaluation:

\begin{itemize}
    \item {Teacher Model Candidates:} DeepSeek-v3, GPT-4o, Claude-3.5-sonnet, Gemini-1.5-pro, DeepSeek-R1, and GPT-o1.
    \item {Student Model Candidates:} DeepSeek-R1-Llama3-8B, Gemma-2-9B-it, Llama3.1-8B-it, Mistral-Nemo-Instruct, Qwen2-2B, Starling-7B, Llama3.2-1B-it, and DeepSeek-R1-Qwen-1.5b.
\end{itemize}

Subsequently, we conducted comparative experiments on two benchmark datasets (described in detail in Section~\ref{sec:experiments}), and the results are presented in Fig.~\ref{fig:various_model_eval}.
The evaluation showed that DeepSeek-v3, GPT-4o, Claude-3.5-Sonnet, DeepSeek-R1, and GPT-o1 consistently outperformed other models across both benchmark sets. Notably, although Claude-3.5-Sonnet achieved an overall accuracy of only 15.3\%, it outperformed DeepSeek-v3 and GPT-4o on the Standard Set, highlighting its strengths in specific scenarios. Based on these findings, we selected these five models as our teacher models.

For the smaller student models, DeepSeek-R1-Llama3-8B, Gemma-2-9B-it, Llama3.1-8B-it, and Mistral-Nemo-Instruct demonstrated clearly superior performance compared to other lightweight alternatives, particularly those with approximately 1 billion parameters (e.g., Llama3.2-1B-it and DeepSeek-R1-Qwen-1.5B). As a result, we selected these four models for further fine-tuning.

After fine-tuning on two distinct training datasets, we produced two sets of fine-tuned student models:

\begin{itemize}
    \item \textbf{HKT-vul Series:} HKT-vul-DeepSeek-R1-llama3-8b-v0.1 (HKT-vul-DeepSeek-R1-llama3), HKT-vul-Gemma-2-9b-v0.1 (HKT-vul-Gemma-2), HKT-vul-Llama3.1-8b-0.2 (HKT-vul - Llama3.1), HKT-vul-Mistral-Nemo-v0.1 (HKT-vul-Mistral).
    \item \textbf{HKT-mix Series:} HKT-mix-DeepSeek-R1-llama3-8b-v0.1 (HKT-mix-DeepSeek-R1-llama3-8b), HKT-mix-Gemma-2-9b-v0.1 (HKT-mix-Gemma-2), HKT-mix-Llama3.1-8b-0.2 (HKT-mix-Llama3.1), HKT-mix-Mistral-Nemo-v0.1 (HKT-mix-Mistral-Nemo).
\end{itemize}

\subsection{Implementation of Fine-tuning Process}
We conducted our model training using an NVIDIA A100 GPU, each equipped with 48GB of VRAM. For optimization, we employed the Adam optimizer with a learning rate of 5e-4. This specific learning rate was selected after extensive experimentation, as it effectively balanced convergence speed and stability for our particular task.
To enhance efficiency and reduce memory requirements, we utilized the QLoRA fine-tuning method instead of LoRA. QLoRA quantizes and compresses the model's weights, allowing for effective fine-tuning with limited computational resources. This approach is particularly advantageous when working with LLMs on hardware with restricted memory capacity, as it reduces the VRAM footprint without compromising the model's performance \cite{zi2023delta}.



\subsection{Implementation Details of RL Evaluation}
Introducing new knowledge into LLMs can inadvertently add noise, potentially reducing the accuracy of fine-tuned models in Section \ref{sec:reinforcement_learning}. To mitigate this issue, we integrate a reward model that computes prediction losses and provides targeted feedback. Inspired by MFTCoder \cite{liu2023mftcoder}, our evaluation approach combines label prediction accuracy with a rationale-based explanation system to promote both correctness and interpretability.

\paragraph{Reward Signal and Training Objective}
We denote the validation dataset (comprising real-world audited contract collections) as 
$\mathcal{D}_{val} = \{(x_i, y_i, r_i)\}_{i=1}^{N}$,
where each \(x_i\) represents an input, \(y_i\) is the corresponding target label, and \(r_i \in \{0,0.5, 1\}\) quantifies the rationale match between the student’s explanation and the audited report( match (1), partial match (0.5), and not match (0)).

The fine-tuned model \(f\) is trained to minimize the label prediction loss, defined using the cross-entropy loss for categorical outputs:
\begin{equation}
  \mathcal{L}_{label} = -\frac{1}{N} \sum_{i=1}^{N} \log p(y_i \mid f(x_i)),
\end{equation}
where \(p(y_i \mid f(x_i))\) is the probability of the model outputting the correct label given the input \(x_i\).

For rationale prediction—a binary classification task—we employ binary cross-entropy loss:
\begin{equation}
  \mathcal{L}_{rationale} = -\frac{1}{N} \sum_{i=1}^{N} \left[ r_i \log g(x_i) + (1-r_i) \log \big(1-g(x_i)\big) \right],
\end{equation}
where \(g(x_i)\) is the model’s predicted probability of rationale presence for input \(x_i\).

The total loss combines these with a trade-off parameter \(\lambda\):  
\begin{equation}
\mathcal{L} = \mathcal{L}_{label} + \lambda \mathcal{L}_{rationale}.
\end{equation}

We set \(\lambda = 0.8\) to prioritize interpretability, with plans to dynamically adjust \(\lambda\) based on calibration metrics (e.g., expected calibration error).

\paragraph{Iterative RL Process}  
The student model is refined via Proximal Policy Optimization (PPO) \cite{schulman2017proximal}, where the reward signal \(R\) at each iteration is derived from:  
\begin{align}
R =\;& \alpha \cdot \text{Accuracy}(y_i, \hat{y}_i) 
     + \beta \cdot \text{RationaleScore}(r_i, \hat{r}_i) \notag \\
   & - \gamma \cdot \text{SafetyViolationPenalty}
\end{align}
with \(\alpha, \beta, \gamma\) controlling the contribution of accuracy, rationale alignment, and rule-based safety penalties. Safety violations (e.g., missing critical vulnerabilities) trigger large negative rewards (\(\gamma = 10\)), enforcing domain constraints.  

\paragraph{Confidence Approximation}
In initial experiments, we simplify confidence estimation:
\begin{itemize}
\item Correct predictions: \(g(x_i) \approx 0.8\) (high confidence),  
\item Incorrect predictions: \(g(x_i) \approx 0.2\) (low confidence).  
\end{itemize}

Under these assumptions, the rationale loss approximates:  
\[
  \mathcal{L}_{rationale} \approx -\frac{1}{N} \left( N_{co} \cdot \log(0.8) + N_{in} \cdot \log(0.2) \right).
\]
where \(N_{co}\) and \(N_{in}\) are counts of correct/incorrect predictions. While this provides a baseline, we later replace fixed values with the model’s actual confidence scores, calibrated via temperature scaling.

\section{Evaluation}
\label{sec:experiments}
In this section, we condut multiple experiments to validate the effectiveness and superiority of our HKT-* series models (latest version) in smart contract auditing.

\subsection{Research Questions}
Our evaluation aims to address the following questions:

\begin{itemize}
  \item \textbf{RQ1: How effectively do the HKT-* series models identify common vulnerability types?} 
  \item \textbf{RQ2: How do the HKT-* series models perform across different top-$N$ approach?} 
  \item \textbf{RQ3: How do HKT-* series models perform on real-world projects datasets?} 
  \item \textbf{RQ4: How effectively do HKT-* series models perform compared to other LLM-based methods?} 
\end{itemize}


\subsection{Experimental Setup}
\subsubsection{Dataset}
To comprehensively measure the model's capabilities, we need robust validation datasets. As mentioned in \cite{zhang2023demystifying}, vulnerabilities in smart contracts can be categorized as either machine-auditable or machine-unauditable. Their survey indicates that existing tools can detect machine-auditable vulnerabilities, with more than 80\% of exploitable bugs falling into this category. Some vulnerabilities are too complex or subtle and require the expertise of multiple human auditors. Thus, for the consideration of detection difficulty, we utilize three validation datasets: Standard Vulnerability set, Real-world Contracts set, and CVE set.

\begin{itemize}
  \item \textbf{Standard Vulnerability Set}: This dataset contains 10 common vulnerability types, which are often targeted by existing vulnerability detection tools. We use the SmartBugs-curated dataset \cite{smartbugs2022}, which is widely used by developers and researchers for analyzing and improving the security of Ethereum smart contracts. It contains 143 annotated contracts with 182 tagged vulnerabilities, each marked with detectable vulnerabilities based on the DASP classification. 
  \item  \textbf{Real-World Projects Set}: While the Standard Vulnerability Set provides relatively simple contracts that are easier for the model to analyze, real-world smart contracts are often far more complex. To evaluate the framework's performance on more intricate and challenging contracts, we use a subset of contracts from the Code4rena-audited projects \cite{durieux2020empirical}, consisting of 72 projects and 6,454 contracts. There are 243 issue contracts and 784 high-severity and mdeium-severity contracts in the dataset, which are used to assess the framework's ability to detect real-world vulnerabilities. 
  \item  \textbf{CVE Set}:  Well-Known smart contract CVEs. As of January 1, 2025, there are 592 smart contract CVEs, predominantly integer overflows. Followed by PropertyGPT \cite{liu2024propertygpt}, we selected 13 representative CVEs: three are integer overflow cases, three involve access control vulnerabilities, four are other logic bug.
\end{itemize}


\newmdenv[
  backgroundcolor=gray!10, 
  linewidth=0.5pt, 
  roundcorner=10pt,
  skipabove=\baselineskip
]{custommdframed}

\subsection{Evaluation Settings}
\subsubsection{Evaluation Criteria}
The primary objective of our evaluation is to assess the effectiveness of the SmartAuditFlow framework in detecting vulnerabilities in smart contracts. We approach this task as a binary classification problem: for each contract, the system must determine whether a specific vulnerability is present or not. To evaluate the framework’s performance effectively, we design domain-appropriate metrics for smart contract vulnerability detection.
The classification outcomes are categorized as follows:

\begin{itemize}
  \item \textbf{True Positive (TP)}: The framework correctly identifies a vulnerability that exists in the contract, as validated against the ground truth.
  \item \textbf{False Positive (FN)}: The framework fails to detect a vulnerability that is present in the contract.
  \item \textbf{False Positive (FP)}: The framework reports a vulnerability that does not exist in the ground truth.
\end{itemize}

\paragraph{Top-$N$ Approach}
We employ the top-$N$ approach to evaluate vulnerability detection. A true vulnerability is considered a TP if it appears within the top $N$ results produced by the model. This method is useful, especially in complex tasks like smart contract vulnerability detection, where vulnerabilities may not always be ranked perfectly. The top-$N$ approach ensures that vulnerabilities detected within the top results still demonstrate model’s effectiveness in identifying critical issues.

\paragraph{Mean Reciprocal Rank (MRR)}
To evaluate the model's ranking quality, we introduce Mean Reciprocal Rank (MRR). This metric measures the rank at which the first relevant vulnerability is found, giving higher scores to models that rank relevant vulnerabilities higher in the list. The MRR formula is:

\begin{equation}
  MRR = \frac{1}{|Q|} \sum_{i=1}^{|Q|} \frac{1}{rank_i}
\end{equation}
where \( |Q| \) is the total number of queries (or contracts) in the evaluation, and \( rank_i \) is the rank of the first relevant result for each query.

\paragraph{Mean Average Precision (MAP): }
In addition to MRR, we use Mean Average Precision (MAP) to evaluate the overall ranking performance. MAP considers both accuracy and the ranking order of vulnerabilities. It is calculated by first determining the Average Precision (AP) for each query and then taking the mean of AP values over all queries. The formula for AP is:

\begin{equation}
  AP = \frac{1}{N} \sum_{n=1}^{N} P(n) \cdot \text{rel}(n)
\end{equation}
where \( P(n) \) is the precision at rank \(n\) (i.e., the fraction of TPs in the top \(n\) results), and \( \text{rel}(n) \) is the relevance of the \(n\)-th result, where 1 indicates a TP and 0 indicates a FN. 

\subsubsection{Implementation Details}
For model validation, we maintained consistency with our training environment by using the same NVIDIA A100 GPUs with 48GB of VRAM. 
To compare the performance of the base models and fine-tuned models, we expanded our evaluation to include three high performance advanced commercial models (DeepSeek-v3, GPT-4o, Claude-3.5-sonnet, DeepSeek-R1, and GPT-o1). These commercial models were integrated into our evaluation pipeline through their respective public APIs. This integration allowed for a seamless comparison between our fine-tuned models and these leading commercial solutions, providing a comprehensive benchmark for our study.

It is worth noting that the evaluation of language models is conducted using a single-attempt (pass@1) approach. In this method, the framework’s ability to identify vulnerabilities is assessed in one pass, reflecting real-world usage where auditors typically rely on a quick, single evaluation rather than multiple attempts. While we acknowledge that LLM outputs can vary due to their inherent probabilistic nature, focusing on a single attempt allows us to evaluate the models in a practical and realistic setting.

\subsection{Experiment Results}
To answer the research questions posed earlier, we present the results of our experiments.

\subsubsection{RQ1-Comparison with other Methods} 

\begin{figure}[ht]
    \centering
    \includegraphics[width=4.5in]{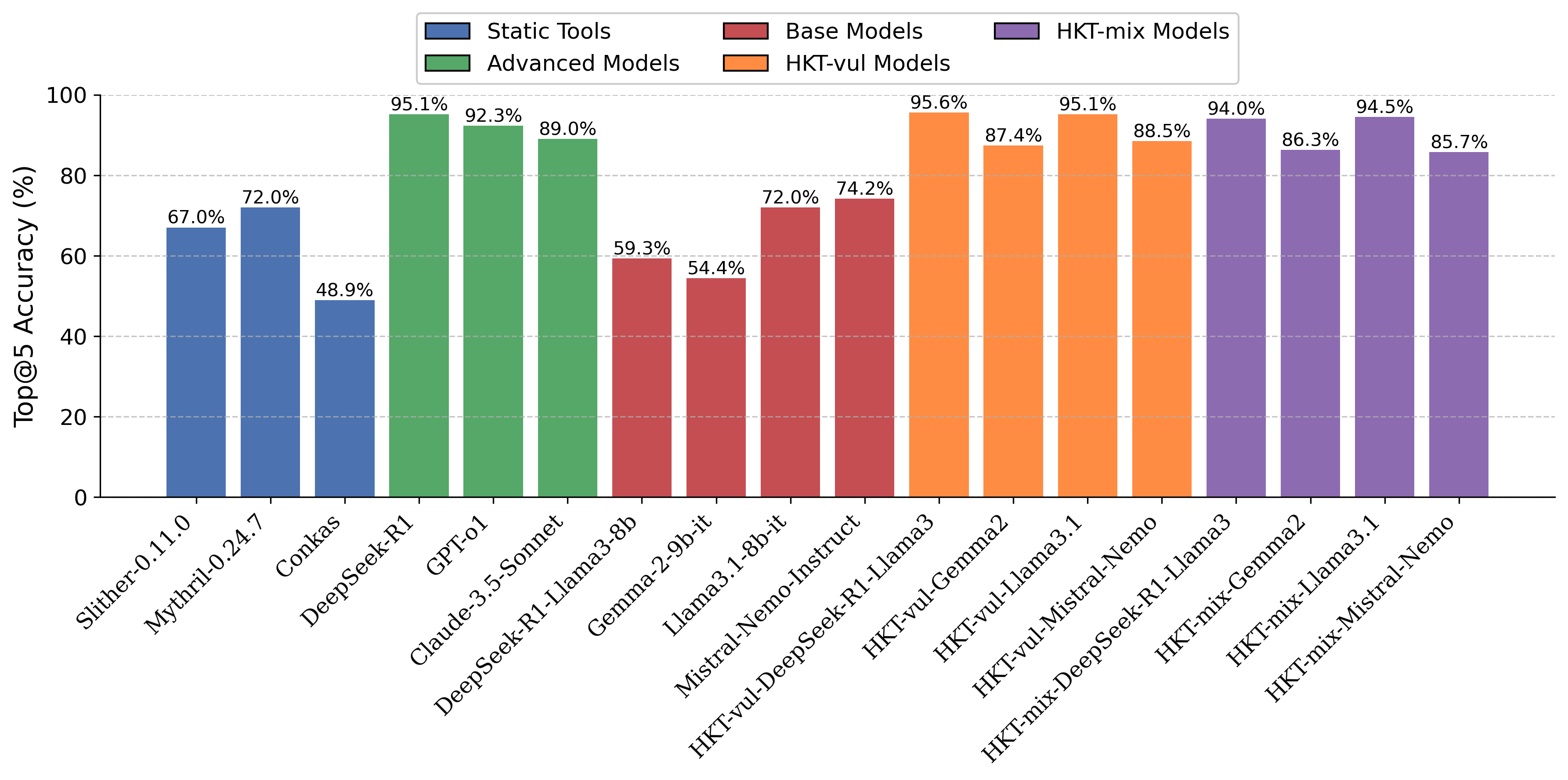}
    \caption{Evaluation of Smart Contract Vulnerability Detection Tools-A Comparative Analysis.}
    \label{fig:stanard_vulnerabiliy_comp}
\end{figure}

To address RQ1, we evaluate the effectiveness of HKT-SmartAudit in enhancing the vulnerability detection capability in smart contracts. Specifically, we focus on detecting 10 distinct vulnerability types defined in the Standard Vulnerability Set and compare the performance of HKT-* series models with several leading traditional static analysis tools, advanced LLM-based models, and their corresponding base models.

For traditional static analysis, we select Slither \cite{feist2019slither}, Mythril \cite{DurieuxFAC20}, and Conkas \cite{di2024evolution}, which have demonstrated superior performance compared to other available options. For advanced models, based on the overall performance discussed in Section \ref{sec:teacher_student_model}, we choose DeepSeek-r1, GPT-o1, and Claude-3.5 for comparison. The evaluation is conducted using a top-max approach. Detailed results are presented in Table \ref{fig:stanard_vulnerabiliy_comp}.

The results indicate that all fine-tuned HKT-* models show significant improvements over their base counterparts. For example, HKT-vul-Llama3.1 achieves 95.1\% accuracy—a 23.1\% improvement over its base model (Llama3.1-8b-it, 72\%). Similarly, HKT-vul-Gemma2 improves by 33\%, increasing from 54.4\% to 87.4\%, while HKT-vul-DeepSeek-R1-Llama3 and HKT-vul-Mistral-Nemo exhibit improvements of 36.3\% and 14.3\%, respectively.

Furthermore, the HKT-* models significantly outperform traditional static analysis tools. For instance, HKT-vul-DeepSeek-R1-Llama3.1 achieves 95.6\% accuracy—surpassing Mythril (72.0\%) by 23.6\% and Slither (67.0\%) by 28.6\%. Even the lowest-performing fine-tuned model, HKT-vul-Gemma2 (84.1\%), outperforms Conkas (48.9\%) by 35.2\%. These results underscore the limitations of rule-based static analysis tools and highlight the potential of AI-driven approaches.

In addition, the HKT-* models not only outperform their base versions and traditional tools but also compare favorably with advanced LLM-based models. For example, HKT-vul-DeepSeek-R1-Llama3.1 achieves slightly higher accuracy than DeepSeek-r1 (an improvement of 0.5\%) and 8.2\% higher than Claude-3.5. These performance gains are particularly noteworthy given that the HKT-* models are significantly smaller and more efficient than their commercial counterparts.

\begin{custommdframed}
  \textbf{Answer to RQ1:} By leveraging HKT-SmartAudit methods, HKT-* series models achieve SOTA performance, outperforming traditional static analysis tools, their base models, and advanced models. Furthermore, the similar performance of the HKT-vul-* and HKT-mix-* variants confirms the consistency and robustness of the proposed approach.
\end{custommdframed}

\begin{table*}[ht]
\centering
\caption{Performance Comparison of Models Using Various Metrics}
\label{tab:model_performance}
\footnotesize
\begin{tabular}{l|ccc|c|c}
\toprule
    \textbf{Model} &  \textbf{top-1} &  \textbf{top-5} &  \textbf{top-max} &  \textbf{MRR} &  \textbf{Average Output} \\
\midrule
    Claude-3.5-Sonnet &  61.0\% &  87.4\% &    89.0\% & 0.71 &     5.9 \\
    DeepSeek-v3 &  47.8\% &  79.1\% &    84.6\% & 0.60 &    15.1 \\
    GPT-4o &  48.9\% &  86.8\% &  91.2\% & 0.65 &   9.0 \\
    DeepSeek-R1 &  33.5\% &  92.9\% &    95.1\% & 0.69 & 6.8 \\
    GPT-o1 &  48.9\% &  90.1\% &    92.3\% & 0.67 &  6.5 \\ \midrule
    DeepSeek-R1-Llama3-8b &  33.5\% &  57.1\% & 59.3\% & 0.41 &  4.5 \\
    Gemma-2-9b-it &  12.6\% &  42.3\% &   54.4\% & 0.32 &   14.5 \\
    Llama3.1-8b-it &  30.8\% &  64.8\% &   72.0\% & 0.49 &    12.1 \\
    Mistral-Nemo-Instruct & 37.9\% &  62.6\% &  74.2\% & 0.39 &  18.1 \\ \midrule
    HKT-vul-DeepSeek-R1-Llama3 &  61.5\% &  \textbf{94.0\%} &  \textbf{95.6\%} & \textbf{0.75} &    \textbf{4.1} \\
    HKT-vul-Gemma2 &  55.5\% &  84.1\% &  87.4\% & 0.59 &  6.1 \\
    HKT-vul-Llama3.1 &  \textbf{66.5\%} &  {81.9\%} &  {95.1\%} & 0.69 &  10.2 \\
    HKT-vul-Mistral-Nemo &  50.6\% &  87.4\% &  88.5\% & 0.61 &  11.1 \\ 
\bottomrule
\end{tabular}
\end{table*}

\subsubsection{RQ2-Evaluation across Different Top-$N$ Approaches}
In addition to the top-max approach used in RQ1, we explore model performance across various top-$N$ thresholds (top-1, top-5, and top-max), along with metrics such as MRR and the average number of outputs generated. These metrics provide insights into both the detection accuracy and the ranking efficiency of each model. In particular, top-1 metric reflects strict prioritization accuracy, top-5 strikes a balance between robustness and practicality, and top-max measures overall detection accuracy. 

Table \ref{tab:model_performance} summarizes the performance of both baseline models and our HKT-SmartAudit variants. A consistent trend across all models is that detection accuracy increases with a larger $N$. However, our HKT-SmartAudit methods not only enhance accuracy across these metrics but also reduce the number of generated outputs, making them more efficient. 
For instance, {HKT-vul-DeepSeek-R1-Llama3.1} achieves the highest top-5 (94\%) and top-max (95.6\%) accuracy, highlighting its superior overall detection capability.
{HKT-vul-DeepSeek-R1-Llama3} ranks first in MRR (0.75) while generating the fewest outputs (4.1 on average), demonstrating its effectiveness in consistently ranking true vulnerabilities early.
Furthermore, the comparison between HKT-vul-Mistral-Nemo and its base counterpart, reveals a significant improvement. HKT-vul-Mistral-Nemo increases top-5 accuracy by 24.8 percentage points (from 62.6\% to 87.4\%) while reducing the average output from 18.1 to 11.1.

\begin{figure*}[t]
  \centering
  \includegraphics[width=0.65\textwidth]{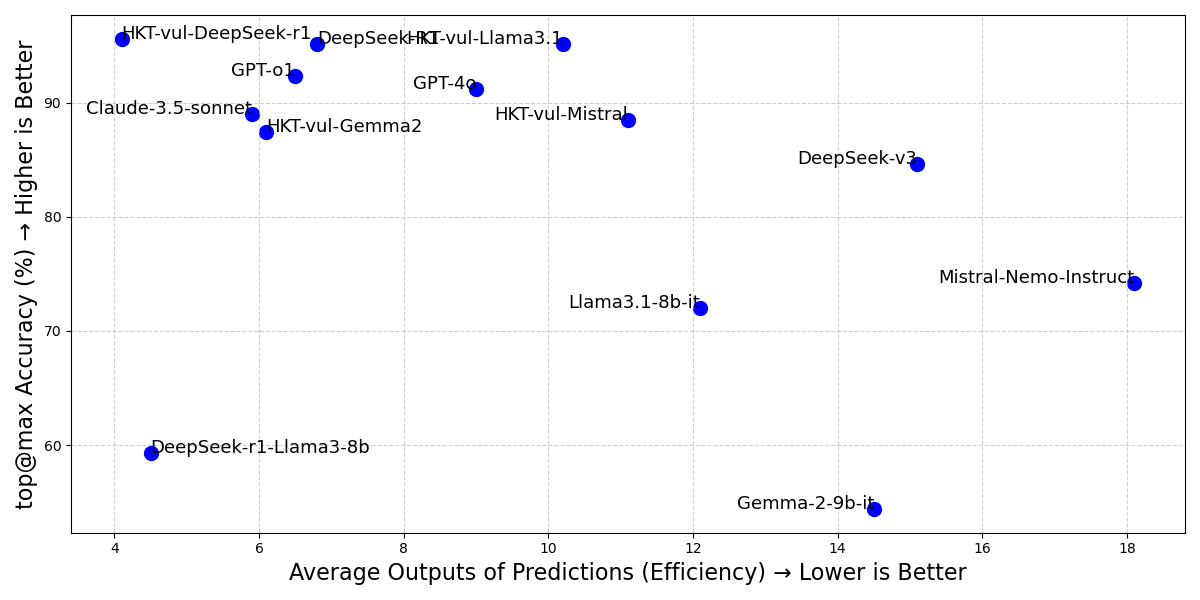}
  \caption{Trade-off Between Accuracy (top-max) and Efficiency on Standard Set.}
  \label{fig:tradeoff_top_accuracy}
\end{figure*}

Fig. \ref{fig:tradeoff_top_accuracy} presents a scatter plot that delineates the balance between top-max accuracy and the computational cost measured by the average number of outputs.  HKT-SmartAudit variants, particularly HKT-vul-DeepSeek-R1-Llama3, occupy the desirable upper left region of the plot. This model achieves a high top-max accuracy of 95.6\% while maintaining minimal outputs (4.1 on average), demonstrating an optimal trade-off between precision and efficiency.
In contrast, other models, such as HKT-vul-Llama3.1, deliver slightly higher accuracy (95.1\% top-max), they do so at the expense of generating more outputs (10.2 on average). This suggests that when accuracy is prioritized over computational efficiency, increased resource consumption is inevitable. 

\begin{custommdframed}
  \textbf{Answer to RQ2:} The experimental results confirm that the HKT-SmartAudit methods significantly enhance detection performance  across different top-$N$ thresholds. By improving both accuracy metrics (top-$N$) and efficiency (MRR), the HKT variants offer a robust and practical solution for real-world smart contract auditing.
\end{custommdframed}

\begin{table*}[ht]
  \centering
  \caption{Evaluation of Smart Contract Vulnerability Detection on Real-World Set}
  \footnotesize
  \label{tab:realworld_performance}
  \begin{threeparttable}
  \begin{tabular}{l |c c c | c | c}
    \toprule
    \textbf{Model} & \textbf{top-1} & \textbf{top-5} & \textbf{top-max} & \textbf{Average Outputs} & \textbf{MAP}\\
    \midrule
    Claude-3.5-Sonnet & 0.6\% & 7.1\% &14.9\%  & 12.5 & 0.123\\
    DeepSeek-v3 & 0.3\% & 11.4\% &21.3\% & 18.8 & 0.148 \\
    GPT-4o & 0.4\% & 13.6\% &21.2\%  & 13.1 & 0.170 \\ 
    DeepSeek-r1 & 1.3\% & 15.7\% &19.9\% & 8.7 & 0.267 \\
    GPT-o1 & 1.1\% & 18.0\% &22.7\% & 7.5 & 0.236 \\ \midrule
    DeepSeek-r1-llama3-8b &0.1\% & 6.3\% &7.7\% & 6.6 & 0.053 \\
    Gemma-2-9b-it & 0 & 1.7\% &6.1\%  & 13.5 &   0.021 \\
    Llama3.1-8b-it & 0.1\% & 3.7\% &9.2\%  & 15.7 &    0.038 \\
    Mistral-Nemo-Instruct & 0 & 1.4\% &4.6\%  & 11.8 &  0.035 \\ \midrule
    HKT-vul-DeepSeek-R1-Llama3 &1.5\% &19.4\% & 24.6\%  & \textbf{6.1} &    \textbf{0.291} \\
    HKT-vul-Gemma2 & 0.3\%&13.9\% & 23.1\%  & 12.1 &  0.171 \\
    HKT-vul-Llama3.1 & 0.4\%&15.4\% & \textbf{25.3\%}  & 11.3 &  0.287 \\
    HKT-vul-Mistral-Nemo & 0.1\% & 16.2\% & 20.3\% & 9.8 &  0.191 \\ 
    \bottomrule
  \end{tabular}
  \end{threeparttable}
\end{table*}

\subsubsection{RQ3-Performance Comparison on Real-World Project Set}
To address RQ3, we perform a comprehensive evaluation of advanced models using the Real-World Project Set, which consists of more complex and challenging smart contracts compared to those used in previous evaluations. Similar to RQ2, this evaluation uses the top-$N$ approach, MRR, and the average number of outputs generated. 
In addition to comparing various models, we also assess traditional tools such as Slither and Mythril. Notably, these tools fail to detect any vulnerabilities in the Real-World Project Set, particularly for logic vulnerabilities \cite{sun2024gptscan}.

The evaluation results are presented in Table \ref{tab:realworld_performance}. The HKT-SmartAudit variants consistently outperform the corresponding base models in accuracy. For instance, HKT-vul-Llama3.1 achieves the highest accuracy at 25.3\%, while HKT-vul-DeepSeek-R1-Llama3 closely follows with 24.6\%. Moreover, HKT-vul-DeepSeek-R1-Llama3 distinguishes itself by generating the fewest average outputs (6.1 per contract), indicating superior efficiency.

Moreover, a clear trend emerges across all models: as the top-$N$ threshold widens, detection accuracy increases. For example, Claude-3.5-Sonnet achieves only 0.6\% accuracy at top-1, which rises to 7.1\% at top-5 and further to 14.9\% at top-max. Similarly, DeepSeek-v3 improves from 0.3\% at top-1 to 11.4\% at top-5 and 21.3\% at top-max. This significant jump from top-1 to top-5 suggests that many models are capable of detecting the correct vulnerability but struggle to rank it as the highest priority. 
In contrast, the HKT variants (e.g., HKT-vul-DeepSeek-R1-Llama3 and HKT-vul-Llama3.1) not only achieve superior top-max accuracies (24.6\% and 25.3\%, respectively) but also show consistent improvements at top-1 and top-5 levels, indicating enhanced detection and ranking capabilities compared to baseline models.

The MAP values further support these findings. For example, HKT-vul-DeepSeek-R1-Llama3 achieves a MAP of 0.291, reflecting its consistent ability to rank true vulnerabilities highly, whereas the base models yield lower MAP values (e.g., GPT-o1 at 0.236 and DeepSeek-r1 at 0.267).

\begin{figure*}[ht]
  \centering
  \includegraphics[width=0.75\textwidth]{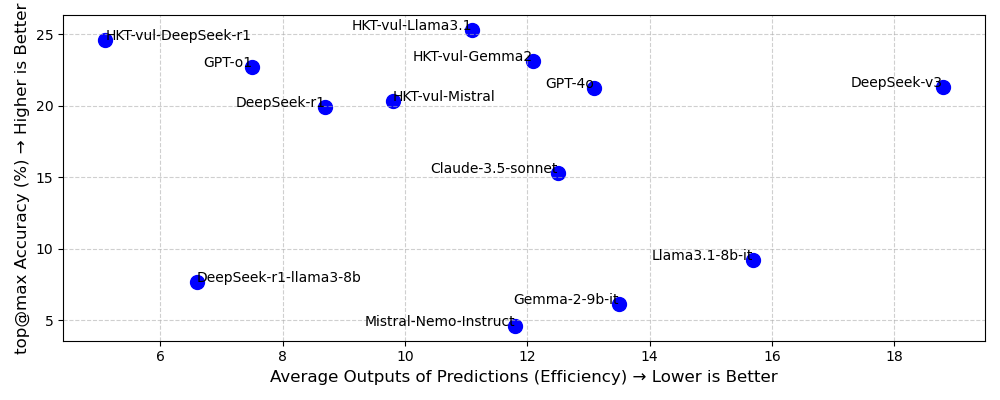}
  \caption{Trade-off Between Accuracy (top-max) and Efficiency on Real-World Project Set.}
  \label{fig:world_tradeoff_top_accuracy}
\end{figure*}

Fig. \ref{fig:world_tradeoff_top_accuracy} presents a scatter plot that illustrates the trade-off between top-max accuracy and computational efficiency (as measured by the average number of outputs). The plot reveals an inverse relationship: models with lower average outputs tend to achieve higher top-max accuracy. In particular, HKT-vul-DeepSeek-R1-Llama3 is positioned in the desirable upper left region of the plot, achieving a top-max accuracy of 24.6\% with only 6.1 average outputs. In contrast, HKT-vul-Llama3.1, despite a marginally higher top-max accuracy of 25.3\%, produces significantly more outputs (11.1 on average).

Furthermore, when compared with common vulnerabilities evaluated in RQ2, logic vulnerability types prove significantly more challenging for language models. This difficulty is evidenced not only by an increase in the average number of outputs but also by a lower detection accuracy.

\begin{custommdframed}
  \textbf{Answer to RQ3:} 
  The experimental results confirm that the HKT-SmartAudit methods significantly enhance detection performance on real-world smart contract projects. By achieving higher top-max accuracy, improved ranking consistency (as reflected by MAP), and greater efficiency (fewer average outputs), our HKT variants offer a robust and practical solution for real-world smart contract auditing.
\end{custommdframed}

\subsubsection{RQ4-Performance Comparison with Other LLM-based Methods}
Recent studies have explored the use of LLMs for smart contract auditing, with notable examples including David et al. \cite{david2023you}, PropertyGPT \cite{liu2024propertygpt}, and GPTScan \cite{sun2024gptscan}. To benchmark our approach against these LLM-based methods as well as traditional auditing tools, we conducted experiments on a CVE dataset. Detailed detection outcomes are presented in Table \ref{tab:cve}, where the detection results for PropertyGPT, GPTScan, Slither, and Mythril were obtained from PropertyGPT’s evaluations.

Our selected model, HKT-vul-DeepSeek-R1-Llama3, correctly identifies vulnerabilities in 10 out of 13 cases, which is comparable to the performance of GPT-o1 (10/13). In contrast, PropertyGPT achieves a detection rate of 9/13, while GPTScan, Slither, and Mythril correctly identify 5, 1, and 3 cases, respectively. These results indicate that our approach and other advanced reasoning models outperform earlier LLM-based methods as well as traditional tools.

Moreover, different methods exhibit varied strengths and limitations. For instance, PropertyGPT fails to detect CVE-2018-14085, CVE-2018-17111, and CVE-2018-17987—vulnerabilities that our model successfully identifies. Conversely, our method misses CVE-2018-18425 and CVE-2023-26488, which are correctly detected by PropertyGPT. These discrepancies suggest that while our approach achieves overall superior performance, each method may have particular strengths for certain vulnerability types.

\begin{custommdframed}
    \textbf{Answer to RQ4:} The experimental results demonstrate that recent reasoning models, including our HKT-* series models, exhibit superior vulnerability detection and reasoning capabilities compared to earlier approaches and traditional auditing tools. Our method, in particular, performs comparably to SOTA reasoning models, confirming the benefits of advancements in model architecture and reasoning techniques.
\end{custommdframed}

\begin{table*}[t]
  \centering
  \caption{Vulnerability detection results for 13 CVEs}
  \scriptsize
  \label{tab:cve}
  \begin{threeparttable}
  \begin{tabular}{l | c | c  c c c  c c c c}
    \toprule
    \textbf{CVE} & Description  & Our work & David & PropertyGPT & GPTScan & Slither & Mythril & GPT-o1 & Deepseek-r1\\
    \midrule
    CVE-2021-34273  & access control  & \checkmark  &  $\times$ & \checkmark & \checkmark & $\times$ & $\times$ & \checkmark & \checkmark \\
    CVE-2021-33403 & overflow & \checkmark & $\times$ & \checkmark & $\times$ & $\times$ & \checkmark & \checkmark & \checkmark \\ 
    CVE-2018-18425 & logic error  & $\times$ & $\times$ & \checkmark & $\times$ & $\times$ & $\times$ & $\times$ & $\times$ \\ 
    CVE-2021-3004  & logic error  & $\times$ & $\times$ & $\times$ & $\times$ & $\times$ & $\times$ & $\times$ & $\times$ \\
    CVE-2018-14085   & delegatecall & \checkmark & \checkmark & $\times$ & $\times$ & \checkmark & $\times$ & \checkmark & \checkmark \\
    CVE-2018-14089   & logic error  & \checkmark & \checkmark & \checkmark & \checkmark & $\times$ &$\times$ &  \checkmark  & $\times$ \\
    CVE-2018-17111   & access control   & \checkmark & \checkmark & $\times$ & $\times$ & $\times$ & $\times$ &  \checkmark & \checkmark  \\
    CVE-2018-17987   & bad randomness   & \checkmark &\checkmark & $\times$ & \checkmark & $\times$ & $\times$ &  \checkmark & \checkmark \\
    CVE-2019-15079   & access control   & \checkmark & $\times$  & \checkmark & $\times$ & $\times$ & $\times$ &  \checkmark  & \checkmark \\
    CVE-2023-26488  & logic error  & $\times$ &  $\times$  & \checkmark & $\times$ & $\times$ & $\times$ & $\times$ &$\times$\\
    CVE-2021-34272   & access control   & \checkmark & $\times$  & \checkmark & \checkmark &$\times$ &$\times$ &  \checkmark & \checkmark \\
    CVE-2021-34270   & overflow  & \checkmark & \checkmark & \checkmark & \checkmark &$\times$ &\checkmark &  \checkmark & \checkmark \\
    CVE-2018-14087  & overflow  & \checkmark & $\times$ & \checkmark & $\times$ & $\times$&\checkmark &  \checkmark & \checkmark \\
    \bottomrule
  \end{tabular}
  \begin{tablenotes}
    \item Note: \checkmark\ indicates a TP, whereas $\times$ indicates a . Our work: HKT-vul-DeepSeek-R1-Llama3.
  \end{tablenotes}
  \end{threeparttable}
\end{table*}


\section{Related Work and Discussion}
\label{sec:related work}
\subsection{Related work}
The application of LLMs in programming has been widely recognized, but their effectiveness in domain-specific languages (DSLs) such as Solidity is an emerging area of interest. Recent studies have begun to explore the potential of LLMs in the domain of smart contract security analysis. 

David et al. \cite{david2023you} explored the application of pre-trained LLMs, including GPT-4 and Claude, for security audits of DeFi smart contracts. Their approach involved a binary classification query, which tasked the LLMs with determining the vulnerability of contracts without additional training. While GPT-4 and Claude showed a high rate of true positives, their use also led to a significant number of false positives. Furthermore, the financial burden of their evaluations was notable, costing approximately 2,000 USD to analyze 52 DeFi contracts. Chen et al. \cite{chen2023chatgpt} conducted a comparative analysis of GPT's performance in detecting smart contract vulnerabilities against other established tools, utilizing a publicly accessible dataset. Their findings revealed that GPT's effectiveness varied across different vulnerability types, focusing on eight common ones, unlike the complex logic bugs in our study.

Similarly, GPTScan \cite{sun2023gpt} tested GPT's ability to match candidate vulnerabilities using a binary response format, where GPT responds with `Yes' or `No' to potential matches with predefined scenarios. This study also pointed out the possibility of false positives due to GPT's inherent limitations. GPTScan analyzed 232 vulnerabilities, correctly identifying 40 true positives. However, although they reported that the advanced GPT-4 model did not yield significant improvements, our research, along with findings from other studies, clearly shows that GPT-4 significantly enhances detection capabilities. Contrary to the previous methods that relied on expensive pre-trained solutions like GPT-4 and Claude, our research leverages advanced fine-tuning of localized LLMs to achieve superior accuracy and cost-efficiency in auditing. 

In addition to commercial products, open-source alternatives have been considered for smart contract analysis. Shou et al. \cite{shou2024llm4fuzz} have integrated the Llama-2 model into the fuzzing process to detect vulnerabilities in smart contracts, aiming to overcome the inefficiencies of traditional fuzzing methods. While innovative, the effectiveness of this approach relies on the accurate and nuanced understanding of smart contracts by LLMs, and it faces challenges in complexity, cost, and reliance on static analysis. Moreover, recent research indicates that traditional methods such as fuzzing often fall short in addressing complex logic bugs. 



\subsection{Summary of Findings}
Based on the above evaluation results, we have derived the following key findings:
\begin{itemize}
    \item \textbf{High Performance of Specialized Models:} HKT-SmartAudit demonstrates that smaller, fine-tuned models can achieve or even surpass the performance of advanced larger models in smart contract auditing. This highlights the benefits of specialized training for domain-specific tasks.
    \item \textbf{Superior Vulnerability Detection and Reasoning:} Our experiments confirm that both advanced reasoning models (e.g., GPT-o1 and DeepSeek-R1) and our HKT-* series models exhibit markedly improved vulnerability detection and reasoning capabilities compared to earlier LLM-based methods and traditional static analysis tools.
    \item \textbf{Efficient Resource Utilization:} The HKT series models are designed for resource-constrained environments, as evidenced by their successful training and deployment on a single NVIDIA Tesla T4 GPU with 16GB memory. This efficiency makes advanced LLMs feasible for practical real-world applications.
    \item \textbf{Balanced Accuracy and Efficiency:} The HKT-SmartAudit approach not only boosts detection accuracy (as seen in various top-$N$ and MRR metrics) but also reduces the number of generated outputs. This indicates an optimal balance between precision and computational efficiency.
     \item \textbf{Extensibility:} While our framework is tailored for smart contract auditing, its design is inherently modular and adaptable, making it extendable to other domains requiring LLM-driven solutions.
\end{itemize}

\subsection{Threats of Validity}
Our proposed system has the following potential limitations:

\begin{itemize}
\item \textbf{Generalization to Unseen Vulnerabilities:}
The training data used in this study is primarily composed of known vulnerability types extracted from publicly available datasets. While this enables targeted learning and evaluation, it may limit the model’s ability to generalize to previously unseen or emerging vulnerabilities that deviate from known patterns. In real-world settings, attackers often introduce novel exploit techniques. Without exposure to such examples during training, the model may struggle to identify them effectively. 

\item \textbf{Knowledge Update:}
Despite strong performance on current datasets, the model’s knowledge remains static post-training. Unlike traditional rule-based systems that can be updated incrementally, LLM-based models require retraining or fine-tuning to incorporate new domain knowledge. This lack of an automated knowledge update mechanism presents a risk of performance degradation over time as new vulnerability patterns, code practices, or language features emerge. 

\item \textbf{Comparison Across Model Sizes:}
Our focus was on demonstrating the effectiveness of smaller, fine-tuned models for smart contract auditing. As such, we did not perform a comprehensive evaluation of larger models within the same architecture families. While our results show that smaller models can outperform some advanced commercial tools, it remains possible that larger models, if fine-tuned similarly, could achieve even higher accuracy. 
\end{itemize}
  
\section{Conclusion}
\label{sec:conclusion}
In this paper, we introduced HKT-SmartAudit, a framework designed to specialize LLMs for the task of smart contract auditing. Our framework integrates four stages of creating specialized models, including data preparation, training process, evaluation, and continuous learning. We showcased the potential and effectiveness of fine-tuning LLMs to achieve high performance in detecting vulnerabilities in smart contracts. Fine-tuned models, especially HKT-DeepSeek-r1-Llama3, exhibit excellent performance in detecting specific types of vulnerabilities, particularly those that are grammar-related or boundary-related. Additionally, we demonstrated that LLMs can be trained to detect complex logic vulnerabilities traditionally identified by human auditors. The quality of the fine-tuning dataset is crucial, as small models, trained on high-quality datasets, can achieve and even surpass the performance of large advanced models. 

In conclusion, HKT-SmartAudit demonstrates the potential of fine-tuned LLMs to significantly enhance the security and reliability of smart contracts, providing a strong foundation for their practical application in real-world scenarios. Our future work will focus on addressing these limitations by enhancing the models' ability to handle complex vulnerabilities and lowering false positives. Additionally, continuous updates and retraining will be necessary to adapt to the dynamic and evolving nature of smart contract development. Moreover, we find that LLMs can not only replace the work of human auditors in identifying security issues but also operate other tools to perform complex attacks on blockchain or smart contracts. Furthermore, LLMs have the potential to fix vulnerable contracts and upgrade them to secure versions.

\bibliographystyle{unsrt}  
\bibliography{references}

\end{document}